\documentclass[preprint2]{aastex}
\newcommand{\dc}{$\delta$~Cir }
\newcommand{\de}{$\delta$~Cir}
\newcommand{\kms}{km~s$^{-1}$ }
\newcommand{\ks}{km~s$^{-1}$}
\newcommand{\nn}{\accent'27A}
\newcommand{\ms}{M$_{\odot}$}
\newcommand{\rs}{R$_{\odot}$}

\newcommand{\m}{.$\!\!^{\rm m}$}
\newcommand{\st}{.\!\!^\circ}

\shorttitle{Three body system $\delta$ Circini}
\shortauthors{P. Mayer, P. Harmanec, H. Sana, \& J.-B. Le Bouquin}

\begin{document}

\title{The three body system $\delta$ Circini}

\author{Pavel Mayer, Petr Harmanec}
\affil{Astronomical Institute of the Charles University, Faculty of Mathematics and Physics,
 V~Hole\v sovi\v ck\'ach 2, 180 00 Praha 8, Czech Republic}

\author{Hugues Sana}
\affil{ESA / Space Telescope Science Institute, San Martin Drive,
Baltimore, MD21218, USA}

\and
\author{Jean-Baptiste Le Bouquin}
\affil{UJF-Grenoble 1/CNRS-INSU, Institut de Plan\'etologie et d'Astrophysique
de Grenoble (IPAG) UMR 5274, Grenoble, France}

\email{mayer@cesnet.cz}

\altaffiltext{1}{Based on data products from observations made with ESO
Telescopes at the La Silla Paranal Observatory under programs ID 65.N-0577, 67.B-0504,
074D-0300, 178.D-0361, 182.D-0356,
083.D-0589, 185.D-0056, 086.D-0997 and 087D-0946.}

\begin{abstract}
\centerline{Version \today}
Delta Cir is known as an O7.5\,III eclipsing and spectroscopic binary with an
eccentric orbit. Penny et al. discovered the presence of a~third
component in the $IUE$ spectra. The eclipsing binary and the third body
revolve around a common centre of gravity with a period of 1644
days in an eccentric orbit with the semimajor axis of ~10~AU.
We demonstrate the presence of the apsidal-line rotation with a
period of $\approx 141$ years, which is considerably longer than its
theoretically predicted value, based on the published radii of the
binary components derived from the Hipparchos $H_{\rm p}$ light curve.
However, our new solution of the same light curve resulted in smaller
radii and a better agreement between the observed and predicted period of
the apsidal-line advance. There are indications that the third body is
a binary. The object was resolved by VLTI with the PIONIER combiner; in
June 2012 the separation was 3.78 mas, magnitude difference in the $H$
region 1\fm75. This result means that (assuming the distance 770~pc) the
inclination of the long orbit is $87\fdg7$.
\end{abstract}


\section{Introduction}

There are not many known eclipsing binaries having such an early
spectral type as \dc (HD~135240, HR~5664, HIP 74778). According to
\citet{walb}, the integrated type is O7.5\,III(f); but see later. The period is
3\fd90, brightness $V=$5\m05 $-$ 5\m20. The orbit is eccentric, so the rotation of
the apside line is to be expected. Knowledge of the apse period is
important, since there is only one other galactic binary of earlier
spectral type with the known apsidal-line rotation period -- HD~93205
\citep{morrell}, and others of similar spectral types as \de, V1007~Sco
\citep{sana,ma2008}, and HD~165052 \citep{052}.

The first three radial velocities (hereafter RVs) of \dc were obtained
at the southern station of the Lick Observatory already in the year 1915
\citep{camp}. Additional RVs were secured by \citet{feast},
\citet{BK65}, \citet{BK69}, and \citet{conti77}. RVs and orbits for \dc
were published by \citet[TE]{thack}, \citet[ST93]{ST93}, and \citet[Pe01]
{Pe01}. The results of these studies were similar; they all found
a small eccentricity of the orbit ($\approx0.06$) and semiamplitude
$K_1\approx150$~\ks.

According to Pe01, the signature of a third body is present in the $IUE$
spectra of \de. Here we analyze the spectra from the ESO archive. The
third body spectrum is present in these spectra, too. The mutual orbit
of the eclipsing binary and the third body is clearly observed.

In the following, we describe the spectroscopic data we used 
(Sect. 2), how the RVs were measured and exploited for the orbital solutions
(Sect. 3), discuss the probable elements of the third body (Sect. 4),
disentangle the spectra (Sect. 5), analyze the interferometric data 
(Sect. 6), present the results of the light-curve solution (Sect. 7) and
discuss the final results (Sect. 8).

\section{The spectroscopic data}
The spectra in the ESO archive consist of 2 UVES
spectra from the years 2000--2001, 29 FEROS spectra from 2007--2009,
and 95 HARPS spectra from 2009--2012\footnote{UVES as well as
FEROS and HARPS are echelle spectrographs. UVES is used with VLT at
Paranal, FEROS \citep{KP} with the 2.2-m ESO/MPI telescope at the ESO
La Silla Observatory and provides spectra in the region from 3625 to
9125~\nn\, with the resolving power of 48000. HARPS is a
spectrograph connected with the 3.6-m ESO telescope, also at La Silla.
Its spectral range is from 3781 to 6911~\nn\, and the resolving power is
115000.}. We downloaded all HARPS and most of FEROS spectra as
pipeline products. We also compiled all available RV measurements from
the astronomical literature\footnote{Two remarks: (1)
While inspecting local RV curves for the 3.9~d orbital period, we noted
that two RVs derived by \citet{BK65} on RJD~37043.2 and 37781.1 deviate
very strongly from the RV curve. There is probably a date error:
increasing both RJD for one day to 37044.2 and 37782.1 brings both RVs
to agreement with the RV curve. (2) In the original paper, \citet{feast}
give a RV = $-84$~\kms for RJD~35252.4, while \citet{thack} tabulate
$-88$~\kms without any explanation. We adopted the original Feast's et
al. value.} and whenever necessary, calculated HJDs for them. Journal
of all available RVs is in Table~\ref{jourv}. Note that throughout this
paper we use an abbreviation RJD = HJD$-$2400000.0\,.

\section{Analysis of radial velocities and orbital solutions}
We first inspected the existing photographic RVs of the primary
component. Several published photographic secondary velocities are
probably rather uncertain. We also used the $IUE$ RVs from Pe01. There
are 41 $IUE$ spectra taken over the course of 17 years. A large part, 29 of them,
was obtained within ten days in September 1992. ST93 used only spectra
from this short interval, and their solution of the primary RV curve has
a rms = 2.3~\ks. Pe01 used all $IUE$ spectra together with the H$\alpha$
RVs (they obtained 18 CAT/CES spectra at the ESO La Silla Observatory
and 3 spectra at the Mount Stromlo Observatory) and got rms = 8.2~\ks.
They found a third component in the $IUE$ spectra. We will show that the
radial velocity of this third body varies; already the rms values
cited above indicate that the position of the third line must vary,
naturally due to the third-body motion around a common centre of gravity
with the 3.9~d binary. Such an explanation of the unexpectedly large rms
has already been put forward by TE. ST93 also noted that their systemic
velocity differs from that derived by TE and explained the difference by
the presence of a third body. Pe01 got different systemic velocities for
the IUE and H$\alpha$ spectra. For the primary RVs, the difference
amounted to 19.6~\ks. Pe01 suggested that this difference might be due to
the fact that the corresponding lines were formed in different layers in
an expanding stellar atmosphere. However, we note that their H$\alpha$
and IUE spectra were obtained in different times, so in view of the
results from the ESO spectra, the true reason of the difference might
well be the orbital motion in the three-body system. In the Pe01
solution, the difference, 19.6~\ks, was added to the H$\alpha$
velocities, so we subtracted this value to use the originally measured
RVs.

The ESO FEROS and HARPS spectra were always taken during rather short
time intervals, every year from 2007 to 2012. The contribution of the
third body to the line spectrum is obviously present in the ESO spectra,
too; it is needed to fill

\begin{deluxetable}{rcrlll}
\tablecaption{Journal of available RVs of $\delta$ Cir\label{jourv}}
\tablewidth{0pt}
\tablehead{
\colhead{Spg.No.}&\colhead{RJD range}&\colhead{No. of RVs}&
\colhead{Observatory/instrument}&\colhead{Source}
}
\startdata
 1 &  20682.65-20740.49& 3/0  &Lick 1-prism   &\citet{camp}\\
 2 &  34522.37-35284.38& 3/0  &Radcliffe      &\citet{feast}    \\
 3 &  36014.00-37781.10& 3/0  &MtStromlo      &\citet{BK65}\\
 4 &  38243.93-38628.87& 4/0  &MtStromlo coude&\citet{BK69}\\
 2 &  39363.20-39906.62&24/5  &Radcliffe      &\citet{thack}\\
 3 &  39703.87-39704.86& 2/0  &MtStromlo      &\citet{BK69}\\
 5 &  42115.81         & 1/0  &Cerro Tollolo  &\citet{conti77}  \\
 6 &  43756.92-49964.99&41/41 &IUE            &\citet{Pe01}     \\
 7 &  49867.54-49874.79&18/14 &ESO coude feed &\citet{Pe01}     \\
 8 &  50152.14-50155.08& 3/3  &MtStromlo CCD  &\citet{Pe01}     \\
 9 &  51653.86-52012.88& 2/2  &ESO UVES       &  This paper     \\
10 &  54277.54-55698.72&29/16 &ESO FEROS      &  This paper     \\
11 &  55003.44-56136.47&95/50 &ESO HARPS      &  This paper     \\
\enddata
\end{deluxetable}

\begin{deluxetable}{lrrrr}
\tablewidth{0pt}
\tablecaption{FEROS and HARPS Radial Velocities of $\delta$~Cir}
\label{CIR}
\tablehead{
\colhead{RJD} &\colhead{Phase\tablenotemark{a}}&\colhead{Primary RV}&
\colhead{Secondary RV}&\colhead{Tertiary RV}}
\startdata
54277.5318 & 0.915 & $  36.6 $ & $  $  \\
54278.5352 & 0.172 & $ 181.4 $ & $-261  $&$-45$ \\
54279.5359 & 0.428 & $ -22.2 $ & $  $  \\
54280.4936 & 0.674 & $-130.9 $ & $288  $ &$-59$ \\
54281.4853 & 0.928 & $  47.5 $ & $  $  \\
54282.4858 & 0.184 & $ 180.3 $ & $-238  $&$-38$ \\
54283.5279 & 0.451 & $ -43.1 $ & $  $  \\
54284.4521 & 0.688 & $-125.5 $ & $274  $ &$-23$ \\
54285.5075 & 0.958 & $  82.0 $ & $  $  &$-49$   \\
54286.5188 & 0.218 & $ 156.9 $ & $-213  $&$-35$ \\
54298.4766 & 0.282 & $ 109.1 $ & $  $  &$-69$   \\
54299.4835 & 0.540 & $-102.3 $ & $239  $ &$-38$ \\
54300.4733 & 0.793 & $ -83.2 $ & $  $  \\
54300.4752 & 0.794 & $ -82.2 $ & $  $  &$-34$   \\
54301.4741 & 0.050 & $ 162.7 $ & $-230  $&$-53$ \\
54302.4761 & 0.306 & $  86.0 $ & $  $  \\
54660.4685 & 0.042 & $ 136.5 $ & $-249  $&$-34$ \\
54660.4707 & 0.042 & $ 135.0 $ & $-250  $&$-31$ \\
54662.6006 & 0.588 & $-139.5 $ & $258  $ &26    \\
54663.5865 & 0.841 & $ -66.0 $ & $  $  &$-5$    \\
54664.5370 & 0.084 & $ 161.3 $ & $-268  $&$-19$ \\
54665.5846 & 0.353 & $  35.3 $ & $  $  \\
54666.4724 & 0.580 & $-137.0 $ & $240  $ &24    \\
54667.4764 & 0.838 & $ -68.3 $ & $  $  &$-4$    \\
54667.5368 & 0.823 & $ -51.1 $ & $  $  &$-5$    \\
54953.7739 & 0.201 & $ 136.1 $ & $-264  $&$-14$ \\
55003.4383 & 0.927 &    15.8 &\\
55003.4419 & 0.928 &    16.9 &\\
55004.4810 & 0.194 &   129.1 & $-268 $\\
55004.4871 & 0.196 &   129.0 & $-269 $\\
55005.7396 & 0.517 & $-116.0 $ &201\\
55006.4428 & 0.697 & $-158.0 $ &265\\
55006.4463 & 0.698 & $-157.4 $ &265 \\
55007.4539 & 0.956 &    45.6 &\\
55008.4769 & 0.218 &   119.9 & $-259 $\\
55009.5571 & 0.493 & $-103.2 $ &  159\\
55010.4657 & 0.728 & $-148.6 $ &249\\
55011.6215 & 0.024 &  105.9 & $-244$\\
55012.4466 & 0.236 &  111.1 & $-244$\\
55028.4563 & 0.338 & $  34.7 $ & $  $  \\
55028.4584 & 0.339 & $  34.3 $ & $  $  \\
55029.4521 & 0.593 & $-147.6 $ & $240  $&70  \\
55029.4547 & 0.594 & $-147.5 $ & $239  $&70  \\
55029.4568 & 0.594 & $-147.9 $ & $240  $&70  \\
55029.5183 & 0.610 & $-153.4 $ & $245  $&60  \\
55029.5204 & 0.611 & $-153.1 $ & $245  $&77  \\
55029.5912 & 0.629 & $-155.3 $ & $245  $&78  \\
55029.6397 & 0.641 & $-153.9 $ & $248  $&72  \\
55029.7290 & 0.664 & $-155.4 $ & $  $  \\
55030.7379 & 0.923 & $  17.6 $ & $  $  \\
55030.7400 & 0.923 & $  17.5 $ & $  $  \\
55030.7428 & 0.924 & $  19.0 $ & $  $  \\
55031.4752 & 0.112 & $ 150.4 $ & $-305 $&$-9$\\
55360.5292 & 0.432 & $ -56.6 $ & $  $ &21 \\
55360.5325 & 0.432 & $ -58.9 $ & $  $  \\
55360.5377 & 0.434 & $ -58.2 $ & $  $  \\
55361.5137 & 0.684 & $-162.3 $ & $231  $&52  \\
55364.4813 & 0.444 & $ -70.6 $ & $  $  \\
55365.4783 & 0.700 & $-155.5 $ & $224  $&27  \\
55365.4825 & 0.701 & $-157.6 $ & $226  $&34  \\
55368.4988 & 0.474 & $ -94.9 $ & $  $ &50  \\
55369.4953 & 0.729 & $-147.7 $ & $210  $&52  \\
55379.4614 & 0.283 & $  71.1 $ & $  $ &12 \\
55380.4548 & 0.537 & $-134.5 $ & $191  $&44  \\
55382.6112 & 0.090 & $ 145.1 $ & $-304 $&22  \\
55383.7235 & 0.375 & $  -8.6 $ & $  $  \\
55736.5051 & 0.775 & $ -79.7 $ & \\
55736.5089 & 0.776 & $ -80.4 $ & \\
55737.5312 & 0.038 & 163.9  &$-206 $&$-43$\\
55737.5371 & 0.039 & 164.0  &$-206 $&$-46$\\
55738.5368 & 0.296 & 107.4  & &$-21$\\
55738.5425 & 0.297 & 105.5  & &$-28$\\
55739.5341 & 0.551 & $ -94.3 $ &258 &$-40$\\
55739.5389 & 0.552 & $ -95.8 $ &258 &$-35$\\
55740.5316 & 0.807 & $ -56.1 $ &203 &\\
55740.5390 & 0.809 & $ -53.7 $ &204 &\\
55741.5105 & 0.058 & 176.2  &$-254 $&$-59$\\
55741.5162 & 0.059 & 177.2  &$-255 $&$-52$\\
55742.4591 & 0.301 & 100.5  & \\
55742.4746 & 0.305 &  98.3  & \\
55743.6485 & 0.605 & $-117.9 $ &294 &$-48$\\
55744.5190 & 0.828 & $ -35.8 $  \\
55744.5257 & 0.830 & $ -34.4 $  \\
55745.4633 & 0.070 & 182.7  & $-260$&$-60$\\
55745.4700 & 0.072 & 184.3  & $-251$\\
55760.4660 & 0.915 &  45.3  & \\
55760.4705 & 0.916 &  50.0  & \\
55760.4901 & 0.921 &  57.8  & \\
55762.4632 & 0.427 & $ -10.8 $ & \\
56103.4443 & 0.803 &$-77.5 $    \\
56103.4500 & 0.804 &$-78.4 $      \\
56104.4482 & 0.060 &156.8 & $-257 $&$-50$   \\
56104.4521 & 0.061 &157.3 & $-262 $&$-50$   \\
56105.4502 & 0.317 & 73.4        \\
56105.4544 & 0.318 & 72.9        \\
56106.4533 & 0.574 &$-117.3 $       \\
56106.4574 & 0.575 &$-121.3 $       \\
56106.4631 & 0.576 &$-119.7 $ &246 &$-28$  \\
56106.4720 & 0.578 &$-120.5 $ &246 &$-22$  \\
56108.4500 & 0.085 &167.6   & $-277 $&$-24$\\
56108.4582 & 0.087 &166.5   & $-278 $&$-32$\\
56109.4479 & 0.341 & 56.7        \\
56109.4525 & 0.342 & 56.7        \\
56109.4575 & 0.344 & 55.5        \\
56110.4490 & 0.598 &$-130.2 $& 265 & $-14$ \\
56110.4563 & 0.600 &$-129.3 $& 278 & $-16$ \\
56111.4608 & 0.857 &$-29.1  $      \\
56111.4675 & 0.859 &$-25.8  $      \\
56113.4493 & 0.366 & 32.5        \\
56113.4562 & 0.368 & 29.9        \\
56132.4625 & 0.238 &134.4  & $-224 $&$-42$ \\
56133.4607 & 0.494 &$-76.9 $ \\
56133.4639 & 0.495 &$-78.4 $ \\
56133.4682 & 0.496 &$-80.8 $ \\
56134.4657 & 0.752 &$-114.7 $&233 &$-26$   \\
56134.4736 & 0.754 &$-112.2 $&233 &$-10$   \\
56135.4592 & 0.006 &114.3 & $-193 $&$-24$  \\
56135.4660 & 0.008 &113.9 & $-194 $&$-24$  \\
56136.4590 & 0.263 &117.4 & $-195 $ \\
56136.4682 & 0.172 &112.8 & $-193 $ \\
\enddata
\tablenotetext{a}{Orbital phases are calculated for the final ephemeris
of Table~3 of the 3\fd9 orbit:
HJD~$2454285.66+3\fd902463\times$E.}
\end{deluxetable}
 the observed He\,{\sc i} line profiles. In
this study, the FEROS as well as HARPS spectra were smoothed to the
resolution 0.06 \nn\, and rectified using the program SPEFO (Horn et
al. 1996, \v Skoda 1996). We also measured their RVs using a fit with
three Gaussians, and -- in the second step -- we applied spectral
disentangling (Sect. 5).

In the He\,{\sc ii} lines 4541 and 5411~\nn\, no traces of the secondary
lines are present. Also no contribution from the third body should be there
since this object is -- according to Pe01 -- cooler than the secondary.
Therefore, we measured the primary RVs in these lines (in
He\,{\sc ii} 4686~\nn\, there is a weak contribution of the secondary).
The secondary RVs were measured in the line He\,{\sc i} 5876~\nn, because
the separation of the secondary from the primary lines at quadratures is
the best here. The secondary velocities were measured only
in cases where their separation enabled a reliable Gaussian fit, i.e.
close to quadratures.

The resulting RVs are listed in Table~\ref{CIR}. In the first 13 HARPS
spectra, a strong fringing was identified by \citet{pore}.
However, we succeeded to find the period and amplitude of the fringes and
to remove them from the spectra.

We derived various orbital solutions from the measured RVs using the
program {\tt FOTEL} \citep{ha04}. When a joint solution for all primary RVs
from the ESO spectra of \dc was derived, the result was quite
disappointing: large deviations were present, with rms 21~\ks. But when
we splitted the available RVs into subsets covering short time intervals
and assigned individual systemic velocity $V\gamma$ to each of them
(which is easy to do with {\tt FOTEL}), a solution with a much lower rms was
obtained. For instance, quite different $V\gamma$ velocities were
obtained for the FEROS spectra from 2007 and from 2008, and for the
HARPS spectra from 2010 and 2011.

\begin{deluxetable}{cccrrcll}
\tablecolumns{7}
\tablecaption{Seasons with $\delta$ Cir Spectral Observations\label{series}}
\tablewidth{0pt}
\tablehead{
\colhead{RJD} & \colhead{Center of} & \colhead{Number of} & \colhead{$V\gamma$} & \colhead{rms} &
\colhead{$\omega$} & \colhead{Source} \\
\colhead{range} & \colhead{the Interval} & \colhead{measurements} & \colhead{\ks} & \colhead{\ks} &
\colhead{deg} & \colhead{of Data}
}
\startdata
39594 to 39723 & 1967.7& 19   &  9.2   & 12.3 &$296\pm17$&TE\tablenotemark{a}\\
44449 to 44460 & 1980.2&  6   &  17.5  &  9.9 &$229\pm10$ &Pe01, $IUE$ RVs\tablenotemark{b}\\
48885 to 48894 & 1992.8& 29   &$ -8.3$ &  4.9 &$257\pm3$&Pe01, $IUE$ RVs\tablenotemark{c}\\
49867 to 49874 & 1995.4& 21   &$ -0.7$ &  9.5 &$272\pm3$&Pe01, H$\alpha$  \\
50152 to 50155 & 1996.2&  3   &$-16.5$ &  8.8 &         &Pe01, H$\alpha$  \\
54277 to 54302 & 2007.6& 16   &  18.6  &  4.6 &$323\pm4$&FEROS 2007 \\
54660 to 54667 & 2008.7&  9   &$ -0.8$ &  4.9 &$305\pm4$&FEROS 2008 \\
55028 to 55031 & 2009.6& 14   &$ -7.9$ &  3.5 &$316\pm4$&HARPS 2009 \\
55360 to 55383 & 2010.5& 13   &$-14.7$ &  2.4 &$310\pm4$&HARPS 2010 \\
55736 to 55762 & 2011.6& 23   &  29.5  &  3.2 &$317\pm3$&HARPS 2011 \\
56103 to 56113 & 2012.5& 21   &  13.6  &  2.7 &$312\pm3$&HARPS 2012/1 \\
56132 to 56136 & 2012.6& 10   &   9.5  &  2.2 &$309\pm3$&HARPS 2012/2 \\
\enddata
\tablenotetext{a}{Only data from the year 1967.}
\tablenotetext{b}{Data from the years 1979 and 1980.}
\tablenotetext{c}{Only data from September 1992.}
\end{deluxetable}

\begin{deluxetable}{lcc}
\tablecaption{The Orbit of the Eclipsing Binary and the Mutual Orbit
with the Third Body based on the Gaussian-fit RVs and a FOTEL
Triple-star Solution}
\label{fot_orb}
\tablewidth{0pt}
\tablehead{
\colhead{Element} & \colhead{3.9~d Orbit} &\colhead{1644~d Orbit}
}
\startdata
Sidereal period $P$ [days] & 3.902463(6)   &1644(3)    \\
Epoch of periastron [RJD]  &54285.66(5) &37482(27) \\
Eccentricity $e$           & 0.0601(48)     &0.415(32) \\
Longitude of periastron $\omega$ [deg] & 308.3(4.7)     & 106(6)\\
Semiamplitudes $K_1$ and $K_3$ [\ks]  & 153.9(1.5) & 23.6  \\
Semiamplitude $K_2$ [\ks]  & 284.17    \\
Systemic velocity $V\gamma$ [\ks]&  3.5     & 3.1   \\
Apsidal advance $\dot\omega$ [deg/day]  & 0.00696(15)\\
Mass ratio $m_2/m_1$       & 0.546(3)  \\
$m_1\sin i^3$ [\ms]          & 22.21\\
$m_2\sin i^3$ [\ms]          & 12.13\\
$a\sin i$   [\rs]          & 33.90\\
\enddata
\end{deluxetable}

\begin{figure}
\resizebox{\hsize}{!}{\includegraphics{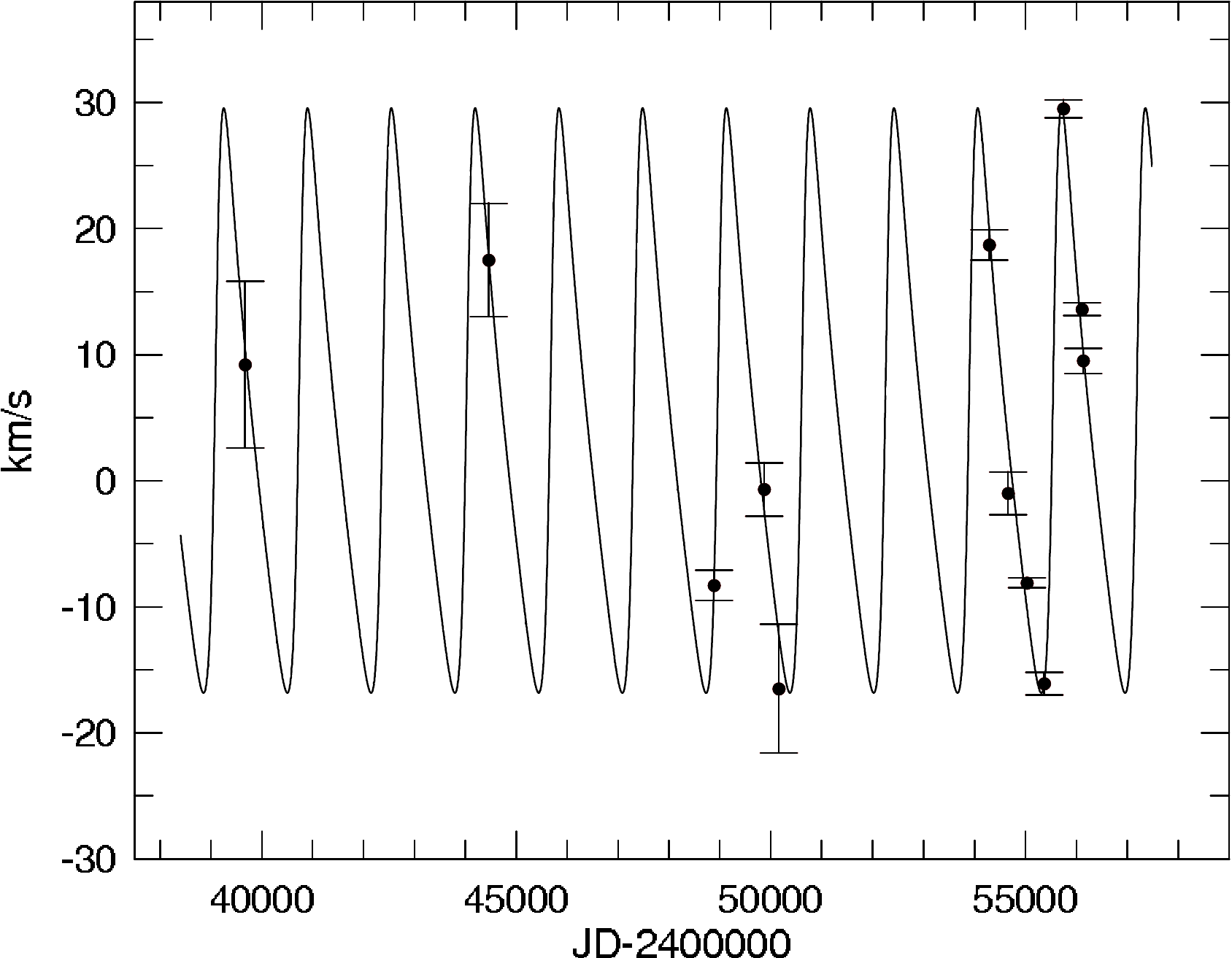}}
\caption{Possible solution of the third body orbit in \de: period 4.5
 years, semiamplitude 23.6~\ks.}
\label{TTT}
\end{figure}

We therefore selected only series of spectra, which cover short time
intervals and derived individual $V\gamma$ velocities for them. Our
selection and the results are listed in Table~\ref{series} and shown in
Fig.~\ref{TTT}. The motion in the mutual orbit is clearly visible. Its
period depends mostly on the ESO data, and it must be close to 4.5
years, with semiamplitude $\approx 24$~\ks. These values fit the older
data too, and both orbits, with the periods of 3.9~d as well as 4.5
year, can be solved. But there is also another way how to use the RVs.
The program {\tt FOTEL} allows to solve both orbits in a hierarchic
system simultaneously, with the advantage that all RVs, not only those
in the time-limited groups, can be used. The solution which includes all
old and new RVs is shown in Table~4. The RVs were weighted
by weights inversely proportial to the squares of rms errors of
individual datasets.

\begin{figure}
\resizebox{\hsize}{!}{\includegraphics{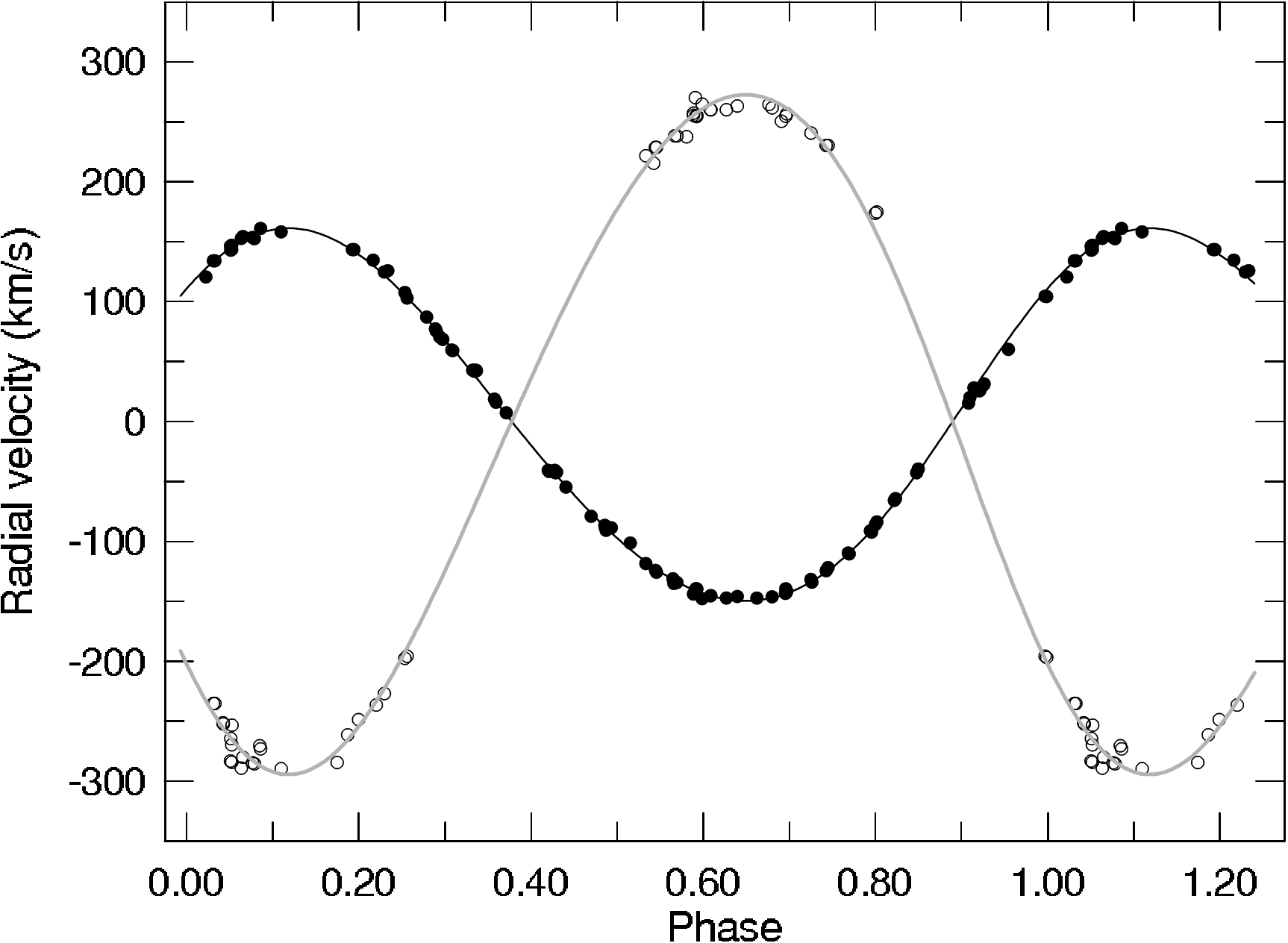}}
\caption{The RVs measured in HARPS spectra corrected to a common systemic
velocity. Full circles -- primary RVs, open circles -- secondary RVs.}
\label{A_H}
\end{figure}

There is certainly a measurable increase of the longitude of the
periastron over the time interval from the $IUE$ spectra to the ESO
spectra. However, the value obtained by TE disagrees with the trend defined
by other sources. This could, of course, be related to the limited number
of photographic RVs available to TE and the effects of the motion in the
wide orbit.

To verify this suspicion, we split the RVs into two data sets, before
RJD~43000 and more recent ones, and derived separate solutions for them
modelling both orbits simultaneously. We obtained periastron argument
values of $\omega=133^\circ\pm25\circ$ for the epoch
RJD~31996 and $\omega=299\st9\pm6\st4$ for the epoch RJD~54286. This would
imply the rate of apsidal advance $\dot\omega=0\st0075$ per day.

\begin{figure}
\resizebox{\hsize}{!}{\includegraphics{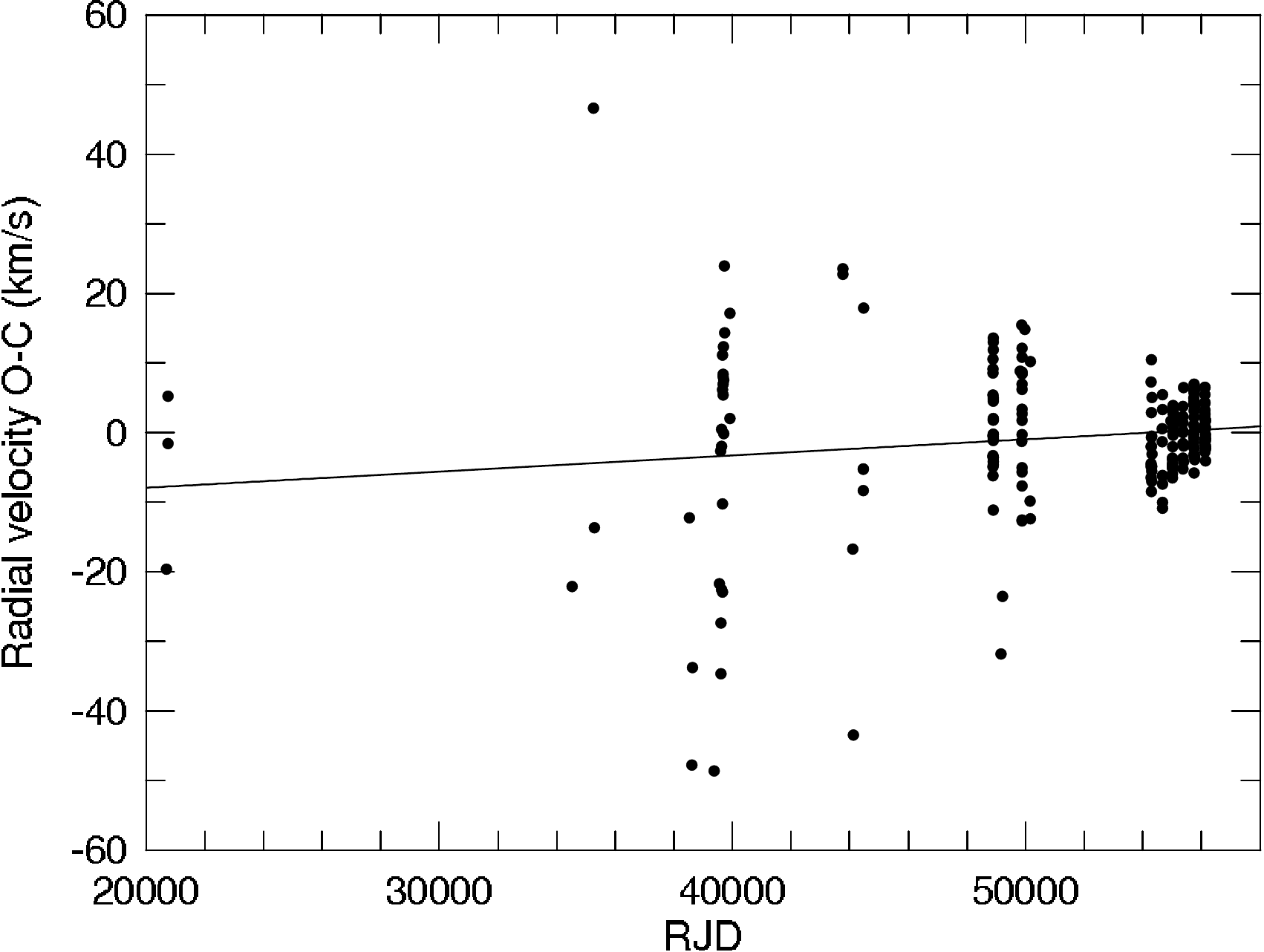}}
\caption{Differences of observed and expected velocities in the third-body
system.}
\label{OMINC}
\end{figure}

We show the measured RVs in Fig.~\ref{A_H}. Due to the apside-line
advance only RVs from a limited time interval -- the RVs measured from
HARPS spectra -- are shown. The RVs are corrected by the systemic
velocity determined for individual seasons (Table~\ref{series}). All
differences of observed and expected velocities in the third-body system
are plotted in Fig.~\ref{OMINC}.

\begin{figure}
\resizebox{\hsize}{!}{\includegraphics{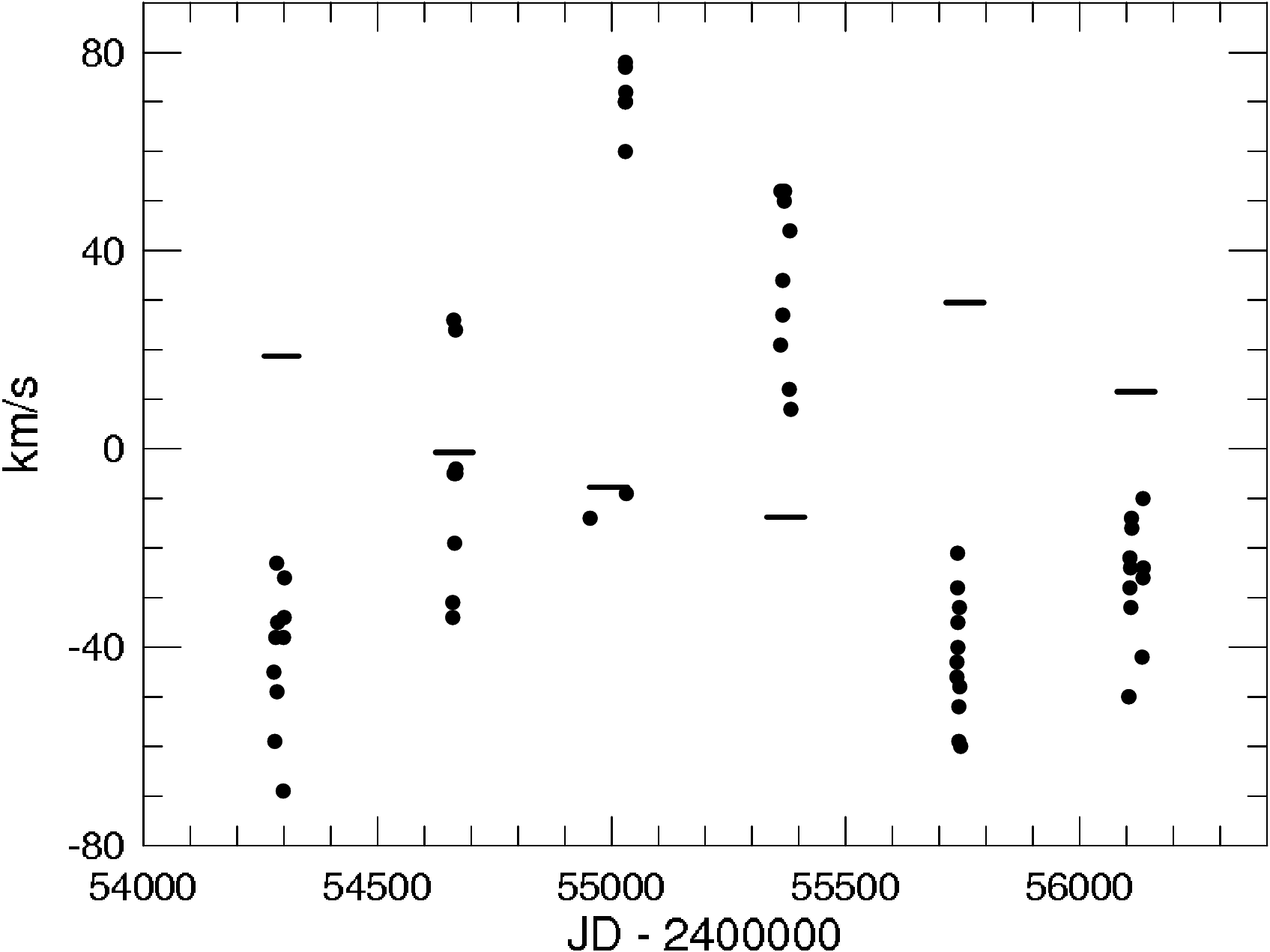}}
\caption{Systemic velocities of the eclipsing binary (short horizontal
 lines) and measured RVs of the third line (full circles) according to
 the FEROS and HARPS spectra.}
\label{SECBIN}
\end{figure}

\begin{figure}
\resizebox{\hsize}{!}{\includegraphics{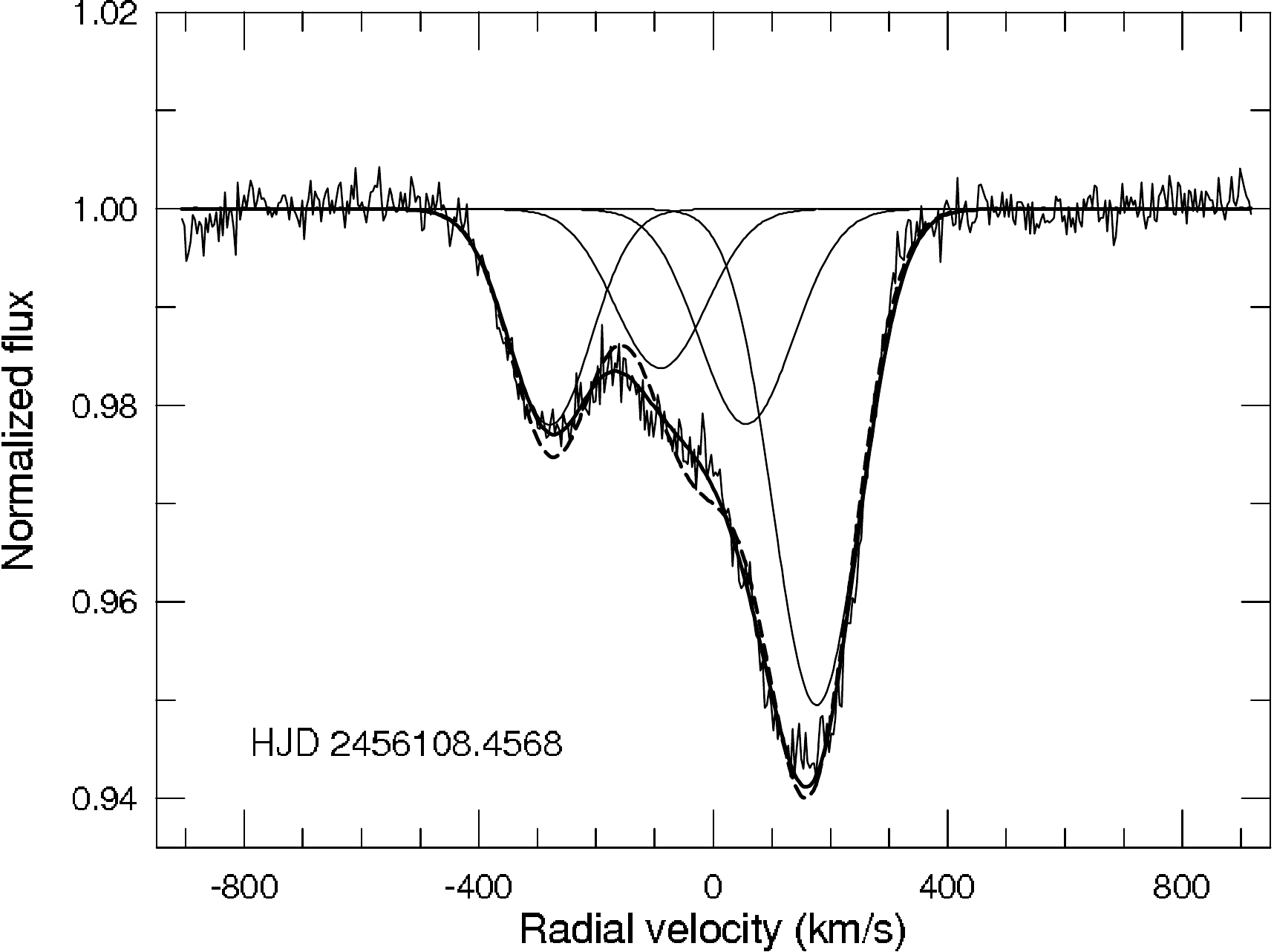}}
\caption{An example of a He\,{\sc i}~4922~\nn\, profile where the two-line fit of
the third-body contribution is neessary. Dashed curve shows the best one-line
fit.}
\label{fourlin}
\end{figure}

\begin{figure}
\resizebox{\hsize}{!}{\includegraphics{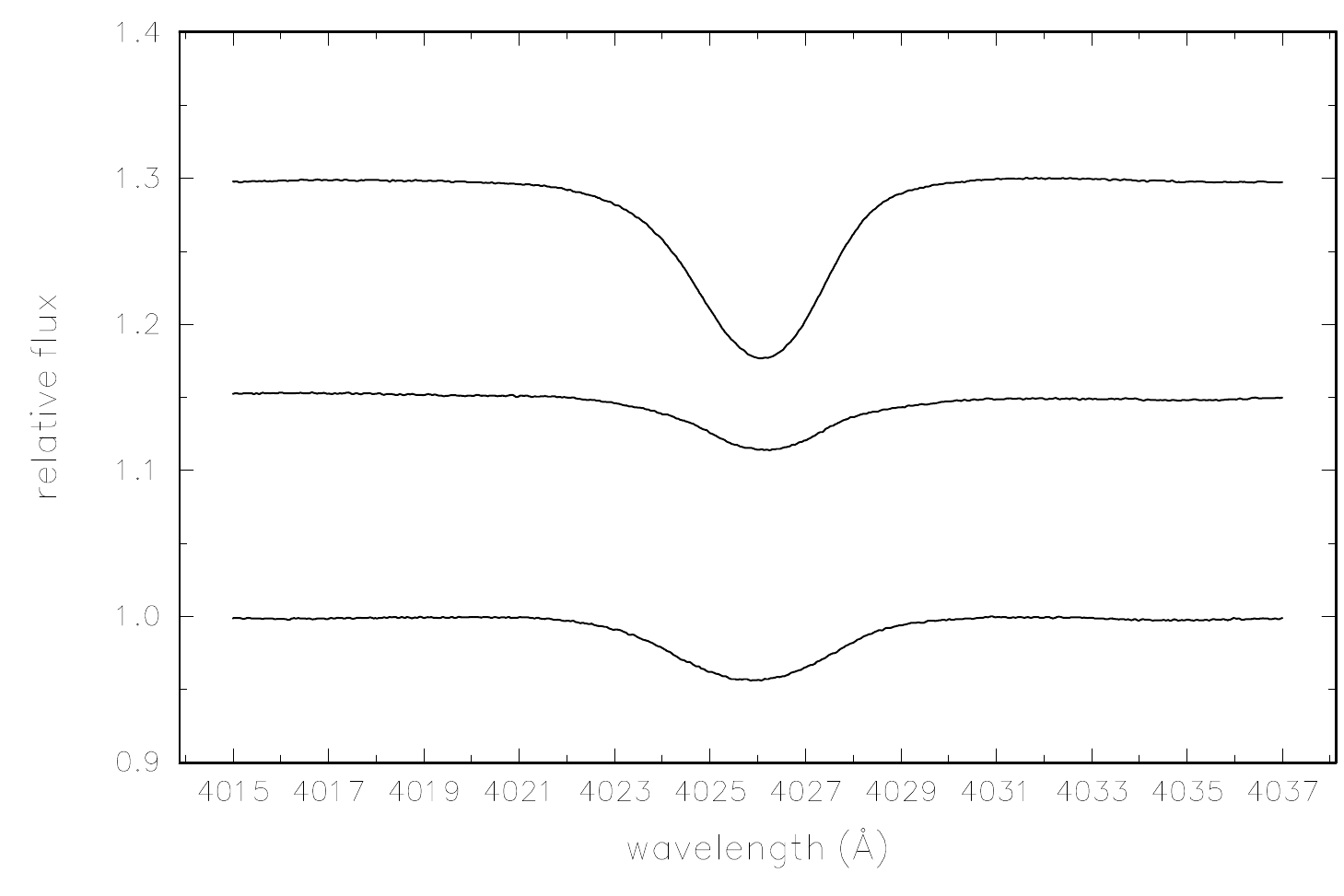}}
\resizebox{\hsize}{!}{\includegraphics{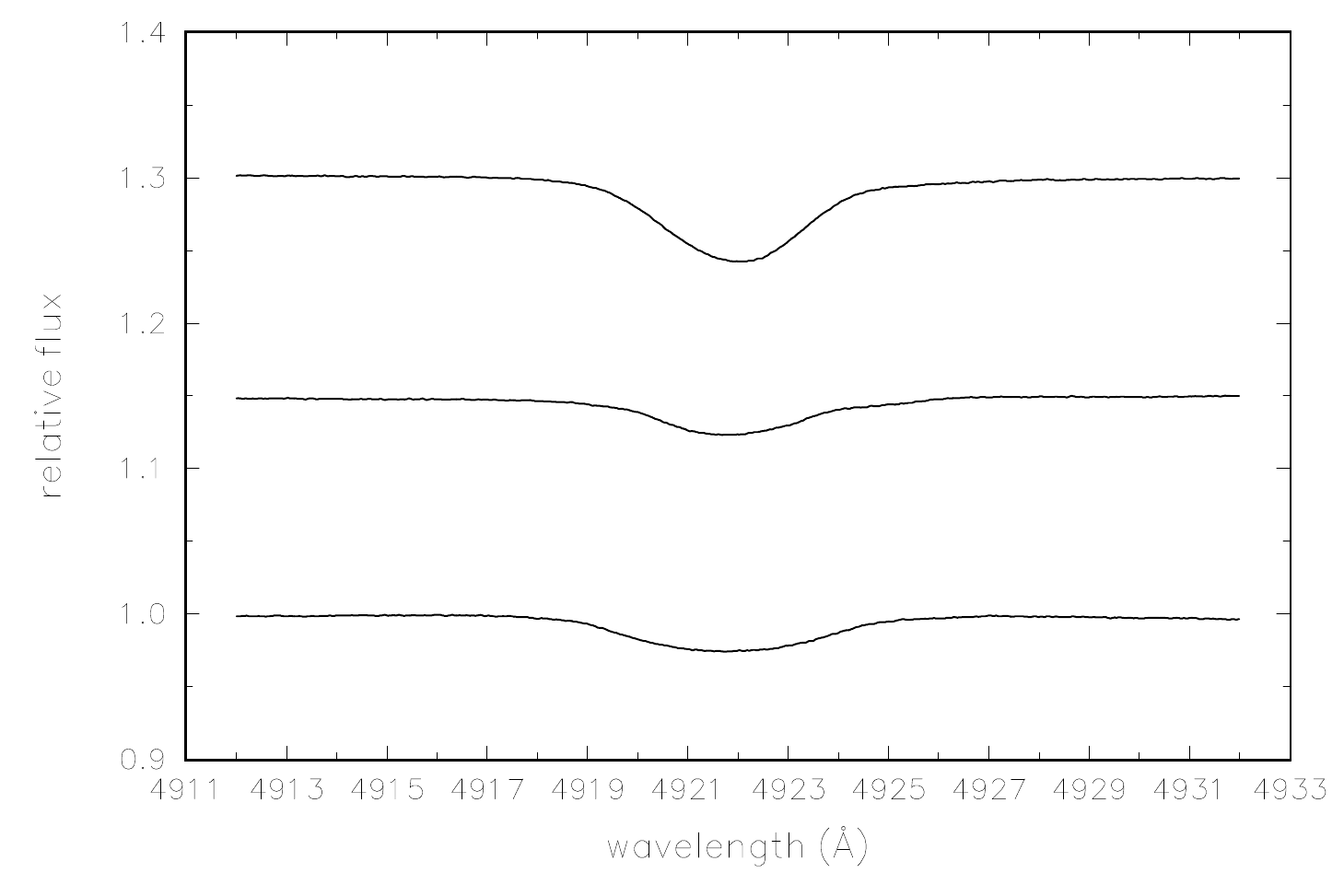}}
\resizebox{\hsize}{!}{\includegraphics{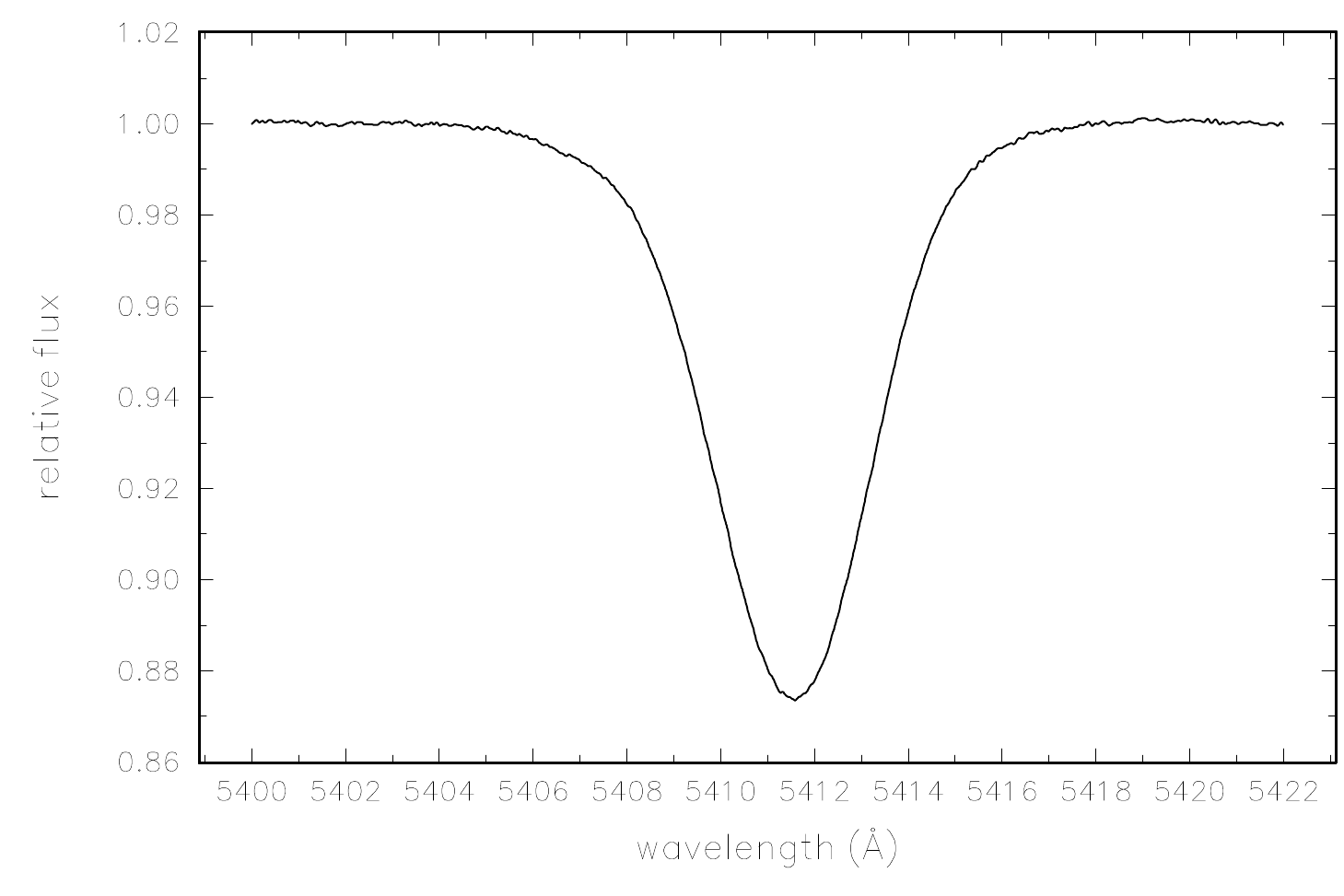}}
\caption{The line profiles of the primary (top profile in each panel),
secondary, and tertiary as disentangled by {\tt KOREL}}.
\label{profKOREL}
\end{figure}

\section{The third body and its mutual orbit with the eclipsing binary}



A relative contribution of the primary is the least pronounced in the
He\,{\sc i} line 4922~\nn. This gives the best chance to measure the line of the
tertiary here, similarly as in the case of the secondary, i.e. near
quadratures. The He\,{\sc i}~4026~\nn\, line, although strongest among the He\,{\sc i} lines,
is not suitable for the Gaussian fitting. For the primary component,
this line is blended with the He\,{\sc ii}~4025~\nn\, line and this would lead
to an incorrectly measured RV of the tertiary. The He\,{\sc i}~4026~\nn\, line is
suitable for disentangling, however, for which the line blending does
not represent any problem.

The systemic velocities of the eclipsing binary (ESO data
only), together with the measured RVs of the third line, are
displayed in Fig.~\ref{SECBIN}. Clearly, the third line velocity varies
in {\sl antiphase} with the systemic velocity, confirming the mutual orbital
motion of both objects. The amplitude of the RV of the third line appears to
be about 2.3 times larger than the amplitude of the orbital motion of the
3.9~d binary in the long orbit as derived from its varying systemic velocity.
Using the program {\tt KOREL} \citep{hb04}, it was possible to estimate
the semiamplitude of the third-body orbit $K_3$ more accurately (see sect.~5
and Table~5).

The third body RVs shown in Fig.~4 were obtained when only a single line
was considered to represent the third body. But it appears that in some
cases a two-line representation of the third light is necessary.
In Fig.~\ref{fourlin} an example of such profile is given. It is usually
difficult, however, to decide which option is better as the fit
qualities of both -- three or four Gaussians -- are quite comparable.
Nevertheless, a possibility that the third body is a binary exists.
Besides the cases where a two-line representation is clearly better,
that possibility is also indicated by the large scatter of all
measurements of the third line RVs. In spite of all effort, we have not
succeeded to find the period of the putative second binary. However,
from the cases where a four-Gaussian fit appears superiour, we suspect
that the period should be of the order of several days.

\section{Disentangling of spectra}

Although the RVs measured using Gaussians gave acceptable results,
more reliable elements can be obtained via spectral
disentangling. We used the program {\tt KOREL} (Hadrava 2004b; details
about the application see Mayer et al. 2013) and applied it to three
different spectral segments, containing the lines He\,{\sc i}~4026,
He\,{\sc i}~4922, and He\,{\sc ii}~5411~\nn. In the case of the
He\,{\sc ii}~5411~\nn\, line, no traces of the secondary or tertiary
components were detected, so only the line of the primary was
disentangled. Since the RVs which were used by the program {\tt FOTEL}
cover longer time interval we assumed the values of both orbital periods
derived by this program (Table~4) when applying {\tt KOREL}.

We first investigated the dependence of the sum of squares of residuals
on various elements, namely the semiamplitudes and mass ratios of both
recognized subsystems, keeping the orbital periods and the value of the
 apsidal advance of the 3.9~d orbit fixed at values obtained
from the {\tt FOTEL} triple-star solution. After finding the values of the
semiamplitudes and mass ratios corresponding to the lowest sum of residuals in
each studied region, we used them as starting values and run a number of
solutions for all elements (but the periods and the apsidal advance), kicking
the initial values to different directions. The solutions with the lowest
sum of residuals are summarized in Table~5 and the corresponding
disentangled line profiles are in Fig.~\ref{profKOREL}.

\begin{deluxetable}{lcccc}
\tablewidth{0pt}
\tablecaption{KOREL Solutions}
\label{ORBITS}
\tablehead{
\colhead{Element}&\colhead{Line 4923~\nn}&\colhead{Line 4026~\nn}&\colhead{Line 5411~\nn}&\colhead{Mean}
 }
\startdata
3.9~d Orbit          &               &             &            \\
\tableline
$T_{\rm periastr.} [RJD]$&54285.612& 54285.634 &54285.672&54285.639 \\
$e$                      &      0.066&      0.072  &      0.067&      0.068 \\
$\omega$ [deg]      &    304.6  &    307.0    &     309.6 &    307.1   \\
$K_1$   [\ks]            &    157.4  &    156.3    &     156.6 &    156.8   \\
$K_1/K_2$                &    0.558  &    0.560    &      --   &    0.559   \\
$K_2$ [\ks]              &    282.1  &    279.1    &      --   &    280.4   \\
$a \sin i$ [\rs]         &     33.84 &     33.51   &      --   &     33.65  \\
\tableline
1644~d Orbit\\
\tableline
$T_{\rm periastr.}$ [RJD]&37469.4  & 37485.0   &  37497.5&37484.0   \\
$e$                      &      0.440&     0.468   &      0.339&      0.416 \\
$\omega$  [deg]          &      284.9&     286.7   &      280.5&      284.0 \\
$K_{1+2}$ [\ks]          &     22.2  &     23.0    &       22.0&     22.4   \\
$K_{1+2}/K_3$            &     0.362 &     0.452   &        -- &     0.407  \\
$K_3$ [\ks]              &     61.3  &     50.9    &        -- &     55.0   \\
$a \sin i$ [AU]          &     11.33 &      9.87   &        -- &     10.64  \\
\enddata
\end{deluxetable}

The values of the mass ratios could not, of course, be derived for the
He\,{\sc ii}~5411~\nn\, region. The equivalent widths (EWs) of several lines as
measured on the disentangled spectra are listed in Table~6.

\begin{deluxetable}{lcccc}
\tablewidth{0pt}
\tablecaption{Equivalent Widths}
\label{EW}
\smallskip
\tablehead{
\colhead{Measurements:}&\colhead{Primary}&\colhead{Secondary}&\colhead{Tertiary}
}
\startdata
EW 4922 \nn &  0.183 & 0.084 & 0.061  \\
EW 4686 \nn &  0.44  & 0.05  & 0.000  \\
\tableline
Theory:\\
\tableline
$T_{\rm eff}$ assumed &  34000 & 29000 & 28000  \\
$\log g$ assumed& 3.75 & 4.1 & 4.2 \\
$L_{\rm i}$    &  0.66  & 0.16  & 0.13   \\
\tableline
EW 4922 \nn      &  0.25  & 0.64  & 0.68   \\
EW 4922/$L_{\rm i}$ &  0.28  & 0.52  & 0.47   \\
\enddata
\end{deluxetable}

As {\tt KOREL} does not provide an estimate of the errors of the
elements, Table~5 contains also the mean values of all three solutions
and their rms errors. While there is a satisfactory agreement of the
values of the majority of orbital elements among the three solutions,
there is a striking difference in the value of the mass ratio of the
outer orbit between the solution based on the He\,{\sc i}~4026~\nn\, and on
He\,{\sc i}~4922~\nn\, lines. In our opinion, this might indicate
that the third body is indeed a binary and that the line blending
between the components of this pair of stars is different for the two
investigated lines. This implies that before the resolution of this
pair of stars, the mass ratio of the 1644~d system must be considered
uncertain.

The plots of disentangled line profiles show that the secondary
He\,{\sc i}~4922~\nn\, line is asymmetric due to the blend with the
O\,{\sc ii} line 4925~\nn. But the tertiary line is symmetric (although
originating in a component/components of similar temperature) and of
different shape. As a matter of fact, its shape is a perfect sinusoid.
Such a profile would result as an integration of the orbital motion
in the putative second binary system. Therefore, it supports our
suspicion that the third body is indeed another binary. Guided by the
opinion reached when these lines were measured using Gaussians, the EW
ratio of the putative second binary lines might be about 2/3, with
similar mass ratio; so the theoretical curve has been calculated as a
sum of two sinusoids. Apparently, also the contribution of the
``secondary" represents the wings of the profile well.

\section{Interferometry}
Independently of the spectroscopic detection, the third body was
discovered by interferometry. \dc was observed in June 2012 (RJD~56090.554)
with VLTI and the PIONIER combiner \citep{LeBou}; the separation was 3.87 mas
and magnitude difference 1\fm75 in the $H$ band.

The observed separation $\rho$ can be calculated as
\begin{eqnarray}
\rho^2 &=& \left({a(1-e^2)}\over{(1+e.\cos \nu)}\right)^2
(1-\sin^2(\omega + \nu).\sin^2i).\nonumber
\end{eqnarray}
With the elements of the 1644~d orbit (Table 5), the true anomaly
$\nu$ equals $150\degr$ for the given date, and the inclination can be
calculated as $i=87\fdg7$. A large inclination is needed to satisfy the
solution of the long orbit as it will be discussed in the next section.

\section{Light-curve solution, physical elements of the system, and the apsidal
advance}
\begin{figure*}
\resizebox{\hsize}{!}{\includegraphics{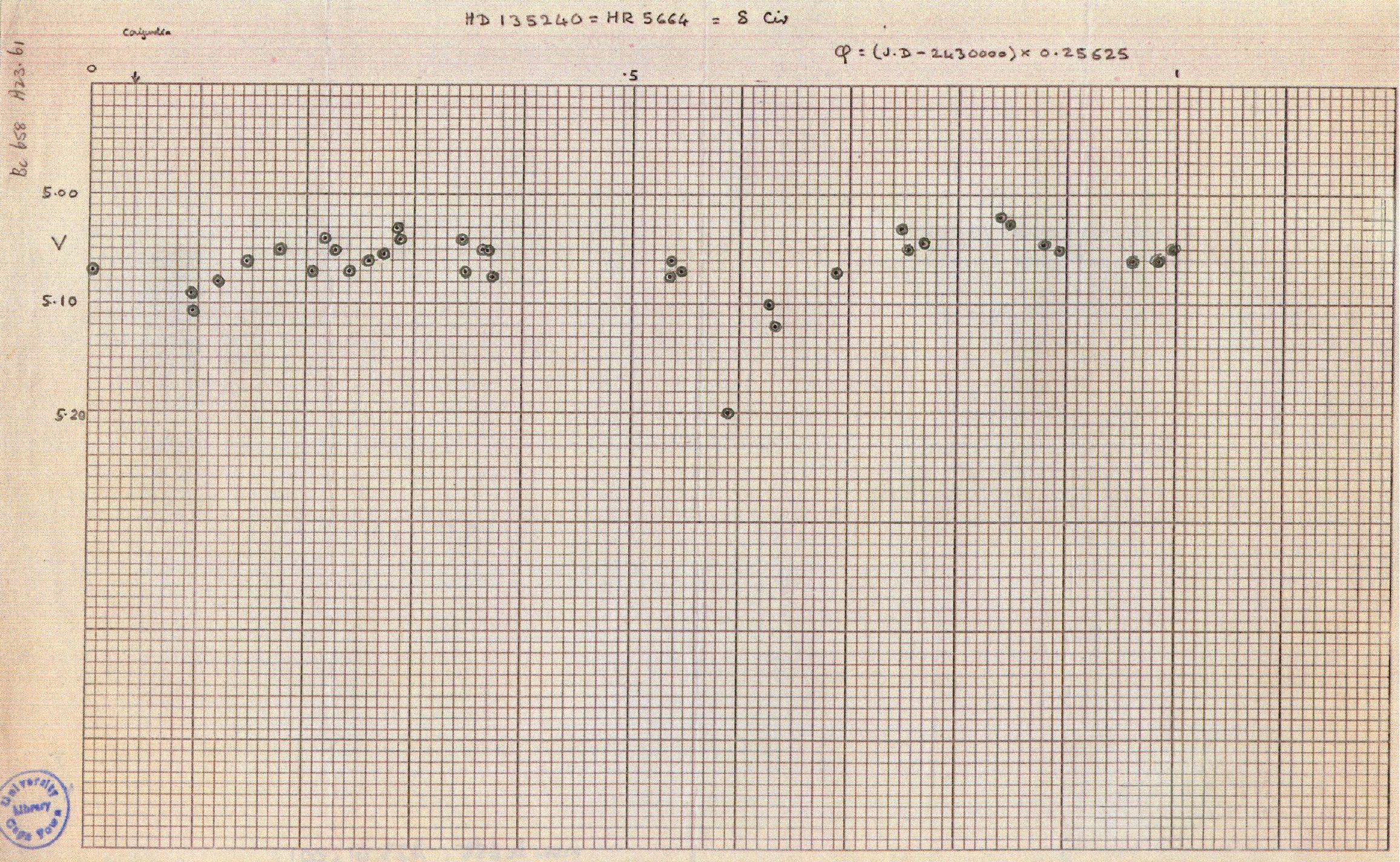}}
\caption{A reproduction of the original phase diagram based on Dr Cousins'
$V$ photometry found in the correspondence of Dr~Thackeray by Dr Kilkenny.}
\label{cousins}
\end{figure*}

The light variability of \dc was first reported by \citet{cou62}. However,
the only published light curve (lc)  of \dc is based on the
{\it Hipparcos} $H_{\rm p}$ photometry \citep{esa97}. We noted
that \citet{thack} mentioned receiving unpublished Cape $V$ photometry of
\dc from Dr Cousins, indicative of binary eclipses. At our
request, Dr David Kilkenny very kindly searched in the archival
materials remaining after the late Drs Cousins and Thackeray.
Regrettably, he was unable to find the original photometric observations
but he did find a plot of the $V$ magnitude vs. phase in the folder with
the correspondence of Dr Thackeray. To preserve this previous piece
of information, we reproduce the original plot in Fig.~\ref{cousins}.
Assuming \citet{thack} ephemeris,
we were able to reconstruct the light curve and use it also along with
the {\it Hipparcos} $H_p$ photometry. In doing so, we took into account
the remark of \citet{thack} that the Cousin's photometry covers an
interval of 5000 days. We therefore assumed that the mean epoch of the
photometry falls into the year 1960 and created artificially fictitious
Julian dates corresponding to the phases of observations for
RJDs 37000-37004.

\begin{figure}
\resizebox{\hsize}{!}{\includegraphics{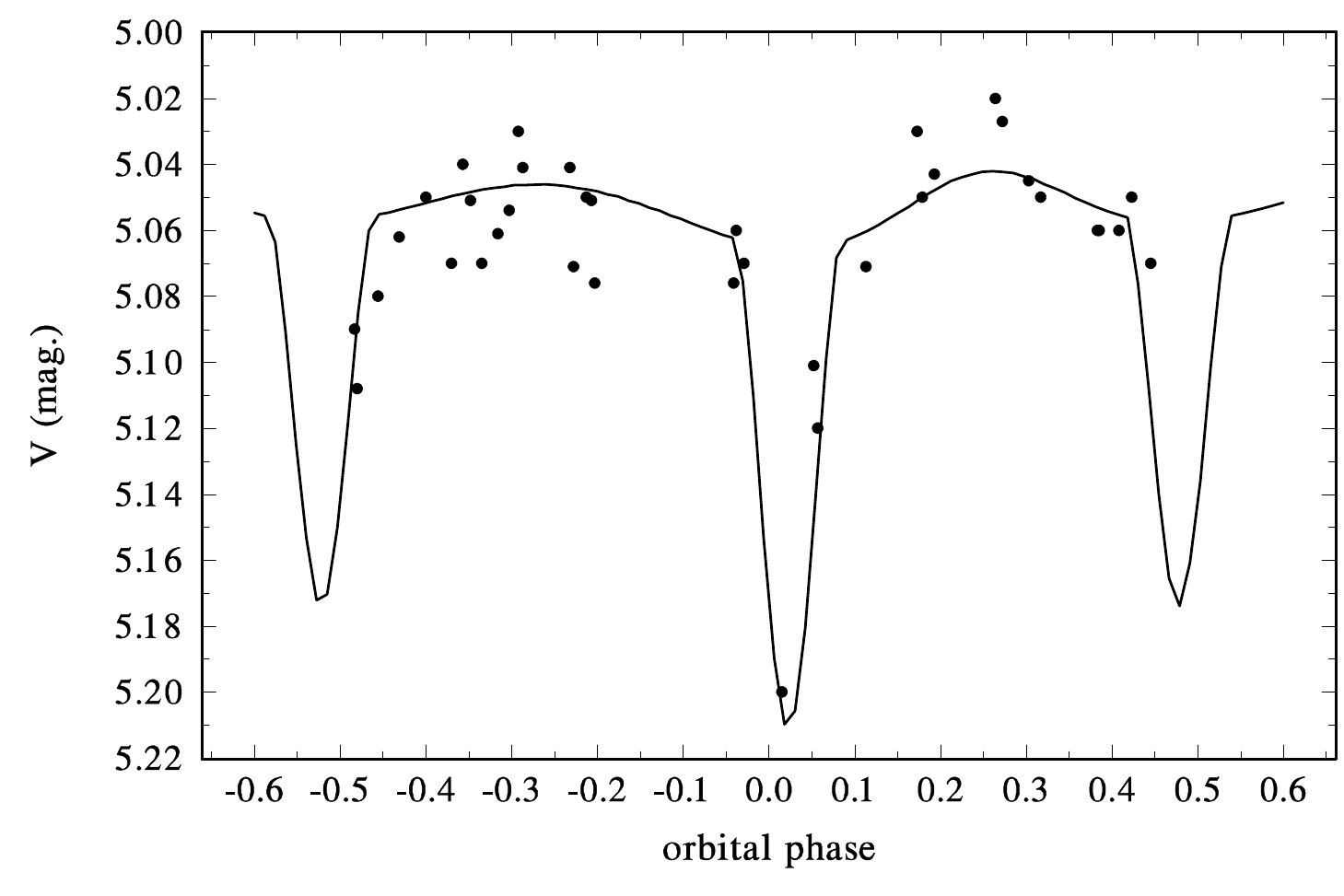}}
\resizebox{\hsize}{!}{\includegraphics{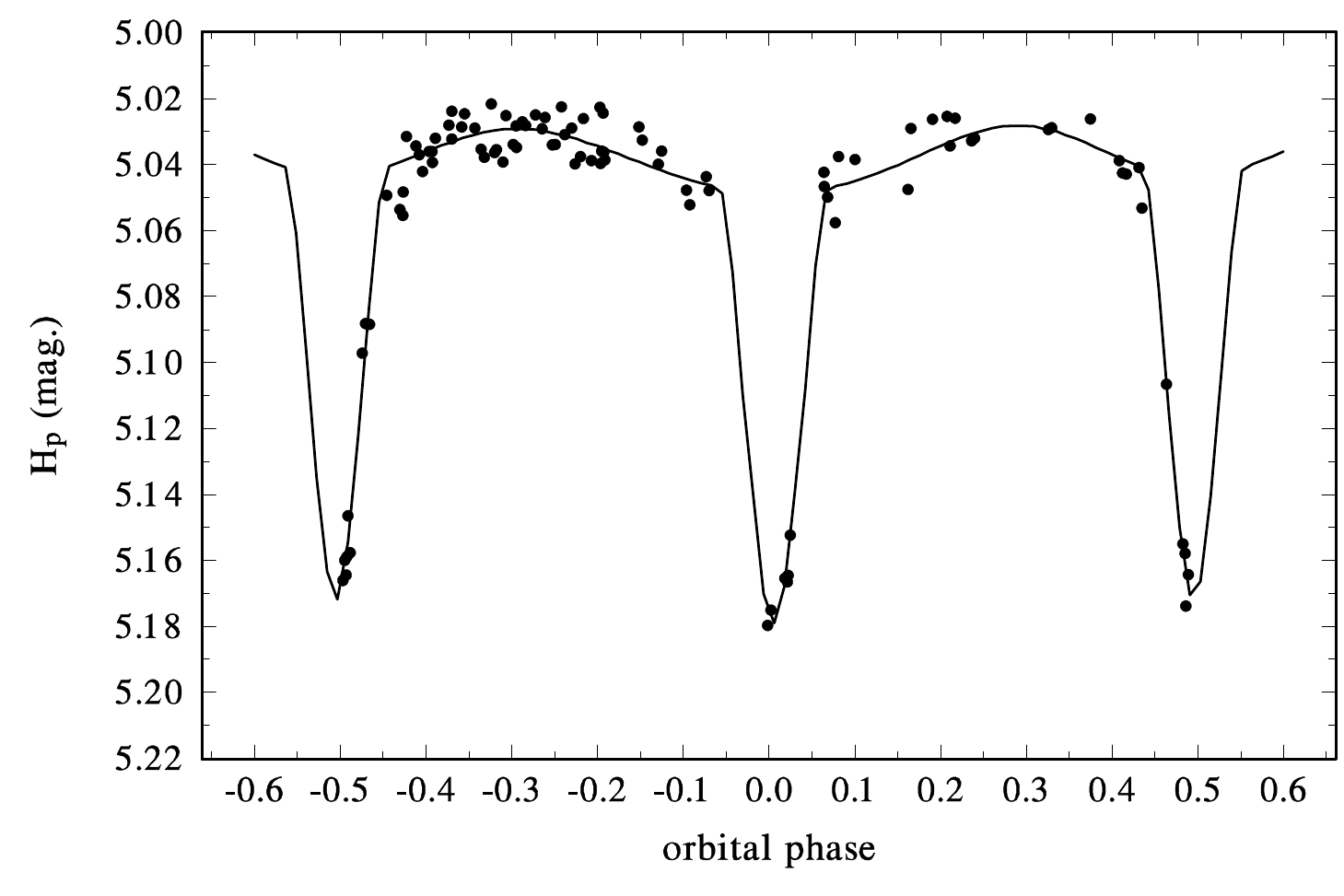}}
\caption{A comparison of observed light curves and {\tt PHOEBE} model light
curves. Upper panel: Cousins Cape $V$ photometry.
Bottom panel: {\tt Hipparcos} $H_{\rm p}$ photometry.}
\label{PHOEBE}
\end{figure}

A solution of the {\it Hipparcos} lc was derived
by Pe01, but they used a program designed for circular orbits only.
It can be seen in Fig.~4 of Pe01 that the observed widths of the
3.9~d binary eclipses are {\sl narrower} than the calculated lc and
the observed maxima are flatter. It was therefore deemed useful
to obtain another light-curve solution, based on a modern program.
We used the program {\tt PHOEBE} \citep{prsa}.
The projected semimajor axis $a\sin i$ and the binary mass ratio
following from the mean {\tt KOREL} solution were fixed in {\tt PHOEBE}.
Since all components of \dc are hot stars we fixed both the albedos and
the gravity darkering coefficients equal to 1. The square-root limb
darkening coefficients were used. The value of the sidereal orbital period
was kept fixed from the final {\tt FOTEL} solution of Table~4.
After a few trials we found that the scarsely covered eclipses cannot
guarantee a unique solution; the space of possible solutions with nearly
identical quality of the fit was immense. To circumwent this unpleasant
situation, we simply assumed the temperatures and relative photometric
radii corrsponding to the known spectral classification of the eclipsing
components, and also the light contibution of the third body as found from
the interferometry and fixed them in the light curve solution.
The only free parameters of the solution were the epoch, longitude
of the periastron passage, orbital inclination, and the relative luminosity
of the primary. Keeping even the inclination of the orbit fixed from
this solution, we then derived an independent solution also for
the reconstructed $V$ light curve from Cousin's observations.
The results are summarized in Table~\ref{lcsol} and the theoretical
and observed light curves are compared in Fig.~\ref{PHOEBE}.




One can see that the fit is very good and in this sense it confirms that
the light curves do not contradict the temperatures and radii we adopted.
More notably, the values of the longitude of the periastron basically
confirm the rate of apsidal advance deduced from spectroscopy.
It is clear, however, that a really accurate determination of
the rate of apsidal advance will require a new, accurate light curve
(and/or continuing spectroscopic observations). At present,
no good photometry regrettably exists. The ASAS data
\citep{poj02} are quite noisy (as it is common for such bright stars),
and only a rough time of a normal minimum can be deduced:
RJD~54601.383 (according to our ephemeris, a secondary minimum should
appear at RJD 54601.381). Any shift of the phase of the secondary
minimum from 0.5 (as it is observed in the {\it Hipparcos} photometry)
is uncertain.

\section{Discussion and conclusions}

From the elements of the mutual orbit as presented in Table~4, the
minimum mass of the eclipsing binary is 43.6~\ms\, and the mass of the
third body 18.7~\ms. However, the total mass of the eclipsing binary
(see sect. 5) is only ~36~\ms. Therefore, $K_3$ is very probably smaller
than 55~\ks; e.g. with 51~\ks\, the minimum mass of the eclipsing
binary is 36~\ms. Certainly the inclination has to be large, which would
be in agreement with the interferometry.

The mass of the third body, as it follows from the solution of the long
orbit, is large (larger than the mass of the secondary) even with the
assumed $K_3=51$~\ks. It would agree better with the interferometric
magnitude difference as well as with the spectroscopic evidence if this
body is a binary.

\begin{deluxetable}{lcccc}
\tablecaption{{\tt PHOEBE} Light Curve Solution and Physical Elements of \de}
\label{lcsol}
\tablewidth{0pt}
\tablehead{
\colhead{Element}&\colhead{Primary}&\colhead{System}&\colhead{Secondary}
}
\startdata
Inclination [deg]      & &$75.81\pm0.10$& \\
Semimajor Axis [\rs]   & &34.7& \\
\tableline
Relative Radii\tablenotemark{a} &0.265 &  & 0.165  \\
$T_{\rm eff}$ [K]      &34000\tablenotemark{a}&  &29000\tablenotemark{a}  \\
Masses [\ms]           & 23.6   & &13.2     \\
Radii  [\rs]           & 9.20   & & 5.73     \\
$M_{\rm bol}$ [mag]    & $-7.76$& & $-6.05$   \\
$\log g$ [cgs]         & 3.88   & & 4.04     \\
$L_{\rm V}$\tablenotemark{b} (Cape) & $0.6460\pm0.0020$  & & 0.1877   \\
$L_{\rm Hp}$\tablenotemark{b} (Hip) & $0.6485\pm0.0005$  & & 0.1852   \\
$L_3$                  & & 0.1663\tablenotemark{a} & \\
$T_{\rm min.I}$  (Cape) [RJD]& &$37003.320\pm0.011$ &\\
$\omega$ (Cape) [deg] &  & $185\pm38$ & \\
$T_{\rm min.I}$  (Hip) [RJD]& &$48398.1212\pm0.0034$ &\\
$\omega$ (Hip) [deg] &  & $256.2\pm2.1$ & \\
\enddata
\tablenotetext{a}{Assumed.}
\tablenotetext{b}{$L_1+L_2+L_3=1$}
\end{deluxetable}

We calculated the expected apside-line period using the evolutionary
models by \citet{claret}. The theoretical period depends strongly on the
radii, and with the radii derived by Pe01 ($R_{\rm pri}=10.2$ and
$R_{\rm sec}=6.4$~R$_{\odot}$; polar radii), the apsidal period should be
$\approx$ 62 years (including the general relativity contribution,
which is 0\fdg00247 per cycle).
In order to reproduce the observed apsidal period, the radii would have
to be smaller. With our radii (Table~7), the corresponding theoretical
apside-rotation period is 85~yrs, closer to but still shorter than the
observed one.

The apside-line period can, of course, be affected by the third body.
From theoretical considerations it is clear that the effect increases if
the ratio of semimajor axis $a_{\rm long}/a_{\rm short}$ is becoming
smaller, and if the mutual inclination increases. In the case of
$\delta$~Cir the ratio of axes is 66, the inclination $11\fdg8<i<16\fdg4$
(since $i_{\rm long}=87\fdg7$ and $i_{\rm short}=75\fdg9$), also $e_1$ is
small. Then the formula by \citet[][eq. 25]{soder84}
might give an approximate idea about the magnitude of the effect (although
it is valid for $e=0$, and $i=0$ only).
According to it, the apsidal period -- due to the third body only -- would
be $\approx 1500$ yrs. As the expected period is $\approx 100$ yrs, the 
effect migh be only of the order of several percents, which is smaller
than the present errors of theoretical estimates as well as of 
observational value.

Once more we repeat that a new, accurate and complete multicolor light
curve is highly desirable and crucial for a reliable determination of the 
basic properties of the binary.

We also note that our new radius of the primary is not consistent
with the radius expected for the O7.5\,III classification, for which e.g.
\citet{martins} give $R\approx~14$~R$_{\odot}$. Another classification
was published by a team from the Yerkes Observatory \citep{hilt}:
O8.5\,V. Pe01 classified the primary component using IUE spectra as
O7\,III-V. We measured the 4471/4541 ratio as $\log W'=+0.21$
\citep{conti71}, the corresponding type is O\,8. Therefore we suspect
that the individual classification of the primary is O8\,IV; such luminosity
class is supported also by the He\,{\sc i}~4143~\nn\, line when compared to
a synthetic spectrum. Pe01 claimed that the primary mass of \dc is smaller
than would correspond to the given spectral type. We got larger
semiamplitudes of RVs, so now the mass is fully consistent with the
value by \citet{martins}.

There are several early B type stars closer than 10 arcmin to \dc
(HD~135160, 135241, 135332), however their distances are not well known
and it is not possible to suggest that \dc\, is a member of a group.

In spite of some uncertainties in the solutions of the RV and light
curves, an acceptable model of the system of \dc is obtained. Further
interferometric data are needed to see if the separation and change of
the position angle follow the prediction.

\acknowledgements
Our very sincere thanks are due to Dr David Kilkenny, who very kindly
searched on several places in the archives to locate Dr Cousins
photometry of $\delta$~Cir and provided us with a phase plot of $V$
observations, which he found.
PM and PH were supported by the grant P209/10/0715 of the Czech Science
Foundation and also from the research program MSM0021620860. We profitted
from the use of the electronic NASA/ADS bibliographical service and
the Strasbourg CDS SIMBAD database.

\end{document}